\documentclass[sigconf,nonacm]{acmart}

\AtBeginDocument{%
  }

\usepackage{booktabs}
\usepackage{array}
\newcommand{\name}{{\textit{TactiPlay}}}

\begin{document}

\title{\name{}: Multi-Granularity Tactical Parsing and Video-Anchored Match Review for Amateur Badminton Players}

\author{Qiaoyi Chen}
\email{qchench@connect.ust.hk}
\affiliation{%
  \institution{The Hong Kong University of Science and Technology}
  \city{Hong Kong}
  \country{China}
}

\author{Yuheng Liu}
\email{liuyh357@mail2.sysu.edu.cn}
\affiliation{%
  \institution{Sun Yat-sen University}
  \city{Zhuhai}
  \state{Guangdong}
  \country{China}
}

\author{Xinzhuang Xiong}
\email{u3584577@connect.hku.hk}
\affiliation{%
  \institution{The University of Hong Kong}
  \city{Hong Kong}
  \country{China}
}

\author{Junze Li}
\email{junze.li@connect.ust.hk}
\affiliation{%
  \institution{The Hong Kong University of Science and Technology}
  \city{Hong Kong}
  \country{China}
}

\author{Hongyi Tang}
\email{hongytang@polyu.edu.hk}
\affiliation{%
  \institution{The Hong Kong Polytechnic University}
  \city{Hong Kong}
  \country{China}
}

\author{Xinyi Zhang}
\email{xzhangfz@cse.ust.hk}
\affiliation{%
  \institution{The Hong Kong University of Science and Technology}
  \city{Hong Kong}
  \country{China}
}

\author{Qingyu Guo}
\email{qingyu.guo@connect.ust.hk}
\affiliation{%
  \institution{The Hong Kong University of Science and Technology}
  \city{Hong Kong}
  \country{China}
}

\author{Xiaojuan Ma}
\email{mxj@cse.ust.hk}
\affiliation{%
  \institution{The Hong Kong University of Science and Technology}
  \city{Hong Kong}
  \country{China}
}

\renewcommand{\shortauthors}{Chen et al.}

\begin{abstract}
Amateur badminton players increasingly record matches, yet existing tools provide only aggregate statistics or generic summaries, leaving most unable to extract tactical insights without expert guidance. A formative study (N=8) reveals the need for multi-granularity, video-anchored tactical analysis centered on rallies. We derive a taxonomy of performance issues from national-level athletes' annotations and present \name{}, an interactive system that instantiates an expert-taxonomy-guided, rally-level, video-anchored review workflow. The system's analytical pipeline organizes match events into taxonomy-grounded feedback, while its interface links structured reports to rally summaries, video evidence, and court visualizations. A within-subjects study (N=16) shows that \name{} elicits more frequent, concrete, actionable, and appropriate reflections than a report-and-statistics baseline. These findings show how organizing reviewed match evidence into a taxonomy-guided, video-linked workflow can support amateur players' tactical reflection.
\end{abstract}




\maketitle

\section{Introduction}

Personal cameras and venue-installed recording services\footnote{\url{https://haoqiuwa.com}} now let amateur badminton players capture and revisit their matches \cite{host2022overview}. When done effectively, retrospective review helps players identify weaknesses, recognize missed opportunities, and track progress \cite{polk2019courttime}. Yet unlike elite athletes, who benefit from professional coaching and structured review workflows \cite{wang2021tacvaluer, wu2022rasipam, wu2021tacticflow, chu2021tivee}, amateurs typically lack an analytical framework. Many cannot spot subtle factors such as positioning inefficiencies, momentum shifts, or poor shot selection, and end up rewatching full matches without deriving actionable insights \cite{zhu2022sporthesia, lin2025sportsbuddy}.

Current AI-powered tools for amateurs, such as Shanyu AI\footnote{\url{https://coachai.net}}, focus on aggregate statistics \cite{wang2020badminton, hsu2024enhancing} or general tips on stroke execution \cite{ma2025t3set}. They seldom deliver event-level tactical reasoning, for instance, articulating how slow recovery led to positional disadvantage and eventual point loss. Such causal reasoning is central to professional coaching \cite{mcgarry2013routledge, peters2013performance} but remains inaccessible to amateurs in existing systems. Bridging this gap requires structured, coach-informed systems that translate video-derived events into multi-level tactical narratives comprehensible to non-expert users.

Two intertwined challenges arise. The first is \textit{analytical}: computer vision can detect rallies, strokes, trajectories, and player positions in racket sports \cite{soloshuttlepose,hsu2024enhancing,tracknet,tracknetv2,tracknetv3,WASB,mmpose,HRnet,monotrack}, but these low-level events do not by themselves provide a structure for tactical reflection. A review workflow must organize events across strokes, tactical sequences, and rallies, map them to coach-informed issue categories, and preserve links to inspectable evidence. Recent work explores LLMs for sensor-to-feedback translation \cite{zhu2025boxingpro, bullard2025enhancing} and VLMs for hierarchical action assessment \cite{wu2025hieroaction}, but how to organize their outputs into an evidence-linked tactical review workflow remains open. The second challenge is \textit{interaction design}: structured feedback alone is insufficient. Players must be able to explore across granularities, from individual strokes through tactical sequences to entire rallies, with each observation grounded in concrete video evidence \cite{sargeant-2015-r2c2, groom-2011-video-pa}.

We address both challenges through an expert-taxonomy-guided, rally-level, video-anchored review workflow. A formative study with eight amateur badminton players informed the post-match review requirements, and annotations from national-level players provided the performance-issue taxonomy. We instantiated the workflow in \name{} through an operational analysis pipeline and an interactive interface for multi-granularity exploration of match evidence. A within-subjects study with 16 amateur players compared this integrated workflow against a report-and-statistics baseline; both conditions used match packages prepared with the same upstream event-detection pipeline, input schema, and manual-verification protocol, and both report generators used GPT-4.1. This work contributes:
\begin{itemize}
  \item A formative study (N=8) identifying design requirements for multi-granularity, video-anchored tactical review, together with an expert-derived taxonomy of performance issues grounded in national-level athletes' annotations.
  \item An expert-taxonomy-guided review workflow, instantiated in \name{} through an analytical pipeline and an interface that connects structured reports, rally-level explanations, match video, and court visualizations.
  \item Empirical findings (N=16) showing that the integrated workflow elicits more frequent, concrete, actionable, and appropriate reflections than a report-and-statistics baseline in a study using reviewed stroke-level and rally-level labels.
\end{itemize}

\section{Related Work}

\subsection{Awareness and Reflective Learning in Sports Pedagogy}

Match awareness, the ability to perceive game situations, evaluate options, and select appropriate actions, is a central goal in sports pedagogy \cite{berry-2008-structured-deliberate}. Game-centered approaches such as Teaching Games for Understanding (TGfU) foreground the cognitive skills of \textit{when} and \textit{why} to apply techniques, not merely their physical execution \cite{bunker-1982-tgfu, cushion-2012-youth-soccer, reid-2007-tennis-skill}. Reflective learning drives this awareness: coaches and athletes use systematic video analysis and guided review to strengthen tactical understanding \cite{grp-reflective-practice-coaches-2024, learning-to-coach-through-experience, knowles-2004-reflective-review, franks-1991-observe-remember, groom-2011-video-pa, pearson-2023-integrative-vfb-pa}.

Sch\"{o}n distinguishes reflection during performance from systematic review afterward \cite{schon2017reflective}. He terms these modes \textit{reflection-in-action} and \textit{reflection-on-action}, respectively. Post-match review operates in the latter mode: players revisit recordings, reframe their understanding, compare observable decisions with rally outcomes, and extract actionable insights. The value lies not in mere replay but in connecting observed events to underlying causes and potential alternatives \cite{gibbs1988learning}.

A gap persists between professional and amateur access to these practices. Professionals rely on advanced coaching systems and structured workflows for deep tactical insight \cite{wang2021tacvaluer, wu2022rasipam, wu2021tacticflow, chu2021tivee}. Amateurs, by contrast, review performances without analytical frameworks and struggle to identify subtle patterns such as positioning inefficiencies and momentum shifts \cite{wang2021tac, lin2025sportsbuddy}. Without structured guidance, their reflection remains informal and yields few actionable takeaways \cite{grp-reflective-practice-coaches-2024, zhu2022sporthesia}. \name{} targets this gap by scaffolding reflection-on-action for amateur players.

\subsection{Video-Based Analysis in Racket Sports}

Video analysis and performance indicators support technical diagnosis and tactical optimization across racket sports \cite{hughes-2002-performance-indicators, liebermann-2002-it-sport, odonoghue-2006-feedback}. Professional systems such as Hawk-Eye track ball trajectories and landing points for in-depth match analysis \cite{fitzpatrick-2024-wimbledon-hawkeye}. Computer vision now automates court detection \cite{monotrack}, player pose estimation \cite{mmpose, HRnet}, shuttle tracking \cite{tracknet, tracknetv2, tracknetv3}, stroke classification \cite{zhu-2006-broadcast-tennis-ar, skublewska-2020-stgcn-tennis, vats-2024-cnnsvm-badminton}, and rally segmentation \cite{see-2025-badminton-rally}. Datasets such as ShuttleSet \cite{shuttleset} and rally outcome prediction models \cite{tan-2022-badminton-rop} enable higher-level tactical inference, while visualization tools like CoachAI \cite{coachai-2019-project} and Sporthesia \cite{zhu2022sporthesia} present these analyses through dashboards and in-video overlays. Closest to our setting, VIRD provides immersive 3D analysis in VR across match, rally, and shot levels for high-performance badminton coaches and players \cite{lin2024vird}. In contrast, \name{} studies an accessible 2D post-match workflow for amateurs, organizing feedback with an expert-derived issue taxonomy and linking it to source video.

Several commercial tools target amateur players directly. In badminton, Shanyu AI\footnote{\url{https://coachai.net}} provides automated recording with static summary reports, and Yuji\footnote{\url{https://apps.apple.com/us/app/yuji-badminton-smash-helper/id6483210189}} focuses on swing recording and form correction. CoachBuddy.ai\footnote{\url{https://coachbuddy.ai}} takes a hardware-driven approach, using IoT-controlled shuttle-launching machines to simulate match rallies and AI-driven opponents. In tennis, SwingVision\footnote{\url{https://www.swingvision.com}} provides on-device line calling, shot tracking, and statistical summaries. Across these tools, feedback commonly takes the form of aggregate statistics, generic summaries, or drill support; direct links from a specific issue to the relevant match evidence and multi-granularity tactical exploration remain limited.

Wearable-based systems take a complementary approach. BadminSense \cite{chen2026badminsense} uses smartwatch vibration signals to classify strokes, predict stroke quality, and estimate shuttle impact location, but does not address rally-level tactical analysis. More recently, LLMs and VLMs have been explored for translating low-level data into coaching feedback. BoxingPro \cite{zhu2025boxingpro} fuses wearable IMU data with LLMs to generate boxing coaching through structured prompts. Iterative prompt engineering has been applied to LLM-based athletic coaching \cite{bullard2025enhancing}. HieroAction \cite{wu2025hieroaction} decomposes action assessment into hierarchical VLM-based sub-action reasoning. Together, these works show how low-level signals can support automated analysis or generated feedback, but they focus primarily on individual actions. They do not address how an amateur player can move across strokes, tactical sequences, and rallies while inspecting taxonomy-guided feedback against the corresponding match evidence.

\subsection{Evidence-Grounded and Interactive Feedback Systems}

Effective feedback requires evidence anchoring and active exploration. Video-anchored feedback can improve understanding and memory, particularly for amateurs who need to inspect concrete evidence \cite{liebermann-2002-it-sport, wulf-2013-attentional-focus, sigrist-2013-augmented-feedback}. Interactive review similarly moves users beyond passive report consumption \cite{baca-2006-rapid-feedback, groom-2011-video-pa}.

YouMove uses augmented-reality mirrors for guided movement comparison \cite{anderson-2013-youmove}, TacticFlow visualizes tactical progressions for post-match reflection \cite{wu2021tacticflow}, and the R2C2 model demonstrates that structured, dialogic reflection improves feedback acceptance \cite{sargeant-2015-r2c2}. Generative AI coaches that produce natural-language advice represent an emerging direction, though reliability concerns persist \cite{joerke-2025-gptcoach}. Together, these systems show the value of linking feedback to visible performance and allowing users to interrogate it. However, they do not establish how to combine taxonomy-guided natural-language feedback, multi-granularity match structure, and direct video verification in an amateur post-match workflow.

\section{Needfinding and Analytical Pipeline}

This section describes the empirical and analytical foundations of \name{} in four steps: a formative study to identify design requirements (\autoref{sec:dr}), expert annotation to derive a performance-issue taxonomy (\autoref{sec:taxonomy}), an operational detection pipeline for extracting match events (\autoref{sec:detection}), and a VLM/LLM-driven stage that maps those events onto the taxonomy to generate structured feedback (\autoref{sec:prepar_system}). The detector components provide event inputs for the review workflow; our contribution centers on how the workflow organizes and links these inputs for reflection.

\subsection{Formative Study}
\label{sec:dr}

We focus on singles matches, the most common recreational format, which allows clearer attribution of tactical choices to individual players. We conducted semi-structured interviews (IRB-approved, $\sim$30 min each) with eight amateur players recruited from a university badminton club.
Participants self-reported Level~2--4 on the \emph{BadmintonCN} nine-level scale\footnote{Level~1 = beginner, Level~9 = retired national-team player. See \url{https://k.sina.cn/article_6434271872_17f833280001004qof.html}.} (distribution: 2, 2, 2.5, 3, 3, 3.5, 4, 4), placing them at the novice-to-lower-intermediate range: they possessed basic stroke techniques, played at least weekly, but lacked systematic tactical coaching. Throughout this paper, \textit{amateur} refers to such recreational-to-lower-intermediate players, as distinct from competitive club players or professionals.

Interviews explored participants' existing reflection practices, the performance aspects they attended to, their assessments of tools such as Yuji\footnote{\url{https://apps.apple.com/us/app/yuji-badminton-smash-helper/id6483210189}} and Shanyu AI\footnote{\url{https://coachai.net}}, and desired features for an ideal review system. Two authors open-coded the interview notes and transcripts, compared the codes, and resolved discrepancies through discussion before consolidating four recurring themes. \autoref{tab:finding-dr} separates each theme from its supporting interview finding and shows the design requirement or requirements it informed.

\begin{table*}[t]
    \centering
    \caption{Four formative themes, the interview findings that define them, and their links to DR1--DR3.}
    \label{tab:finding-dr}
    \small
    \renewcommand{\arraystretch}{1.08}
    \begin{tabular}{@{}>{\raggedright\arraybackslash}p{0.24\textwidth} >{\raggedright\arraybackslash}p{0.52\textwidth} >{\raggedright\arraybackslash}p{0.16\textwidth}@{}}
        \toprule
        \textbf{Formative theme} & \textbf{Supporting interview finding} & \textbf{Linked requirement(s)} \\
        \midrule
        \textbf{Vague impressions without a structured reflection framework}
        & Players often relied on broad post-match impressions rather than a structured way to examine causes; aggregate summaries alone did not help them connect those impressions to specific moments in their own footage.
        & \textbf{DR1}\newline Video-linked evidence \\
        \addlinespace[2pt]
        \textbf{Need for feedback grounded in concrete video evidence}
        & Players wanted each identified issue tied to observable match evidence. Execution problems could be inspected at a single stroke, whereas positioning or recovery problems required the surrounding action--response--consequence sequence to interpret the play in tactical context.
        & \textbf{DR1, DR2}\newline Video-linked evidence;\newline tactical context \\
        \addlinespace[2pt]
        \textbf{Limits of stroke-only feedback for tactical interpretation}
        & A technically acceptable stroke could still lead to an ineffective follow-up or reflect an inappropriate shot choice; stroke-focused tools also missed attacking opportunities and ineffective combinations.
        & \textbf{DR2}\newline Tactical context \\
        \addlinespace[2pt]
        \textbf{Rally as the natural unit of reflection}
        & Players treated rally outcomes as natural starting points and wanted to trace wins and losses to decisive terminal strokes and both players' positions at key moments.
        & \textbf{DR3}\newline Rally-level entry \\
        \bottomrule
    \end{tabular}
    \Description{A three-column table mapping four formative themes and their supporting interview findings to three design requirements. Vague impressions without a structured reflection framework inform DR1. The need for concrete video evidence, including exact strokes or short action-response-consequence sequences, informs DR1 and DR2. The limits of stroke-only feedback for tactical interpretation inform DR2. Treating the rally as the natural unit of reflection informs DR3.}
\end{table*}

Together, these themes yielded three design requirements. \textbf{DR1: Feedback linked to concrete match footage.} Each identified issue should be tied to the relevant moment or short action sequence in the player's own match video. \textbf{DR2: Tactical awareness, not just stroke execution.} Feedback should address follow-up actions, shot choices, missed attacking opportunities, and ineffective shot combinations as they unfold in context. \textbf{DR3: Rally-level outcomes as entry points.} Feedback should summarize decisive terminal strokes and visualize both players' positions at key moments, connecting local technical and tactical problems to why a rally was won or lost. These requirements guided the design of \name{}; the following sections describe how they were operationalized in the review workflow.

\subsection{Expert Annotation and Taxonomy Construction}
\label{sec:taxonomy}

To ground model-generated feedback in professional expertise, three national-level badminton players (each with $>$10 years of competitive experience and coaching involvement) annotated seven amateur match videos ($\sim$8 min each, fixed rear-court viewpoint, players rated Level~2--4). The experts provided free-form, stroke-by-stroke commentary focusing on technical execution and tactical awareness, tracing how the preceding two shots influenced the current action. This produced 304 annotated issues ($\sim$43 per video).

\begin{figure*}[htbp]
    \centering
    \includegraphics[width=\linewidth]{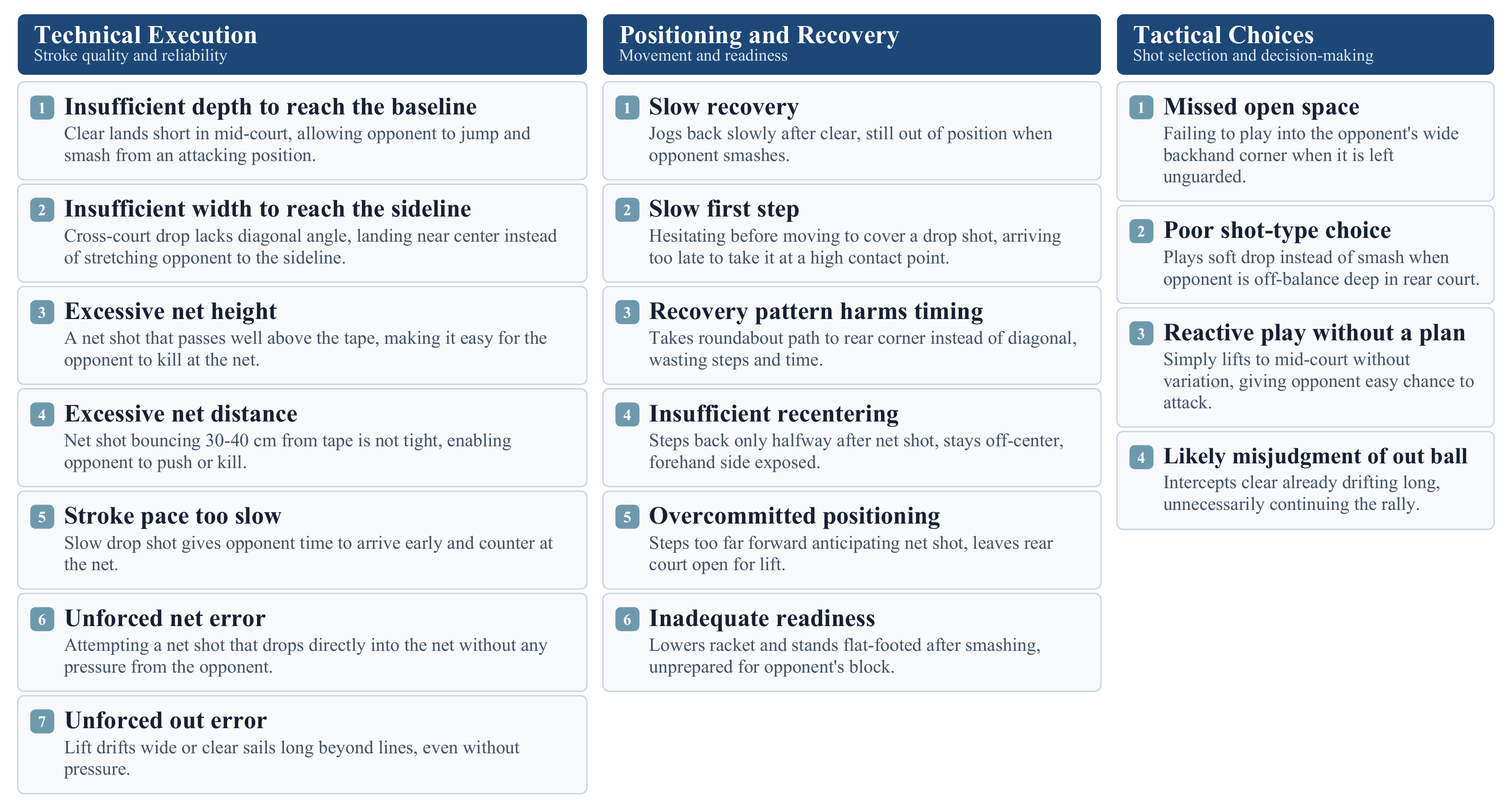}
    \caption{Expert-derived taxonomy of amateur badminton performance issues. Open coding of 304 annotations from national-level players produced three domains with 17 sub-problem descriptors and their operational definitions.}
    \Description{A three-column taxonomy of 17 amateur badminton performance issue descriptors. Technical Execution contains seven descriptors about stroke depth, width, net clearance, pace, and unforced errors. Positioning and Recovery contains six descriptors about recovery speed, first-step timing, recentering, positioning, and readiness. Tactical Choices contains four descriptors about open-space use, shot selection, reactive play, and out-ball judgment. Each descriptor includes a one-sentence definition.}
    \label{fig:taxonomy}
\end{figure*}

Two researchers applied open coding to these annotations, iteratively resolving discrepancies to produce a hierarchical taxonomy (\autoref{fig:taxonomy}) with three domains: \textit{Technical Execution}, \textit{Positioning and Recovery}, and \textit{Tactical Choices}, each with sub-problem descriptors. This taxonomy provides the structured reference for model-generated feedback.

\subsection{Video-to-Report Analysis Pipeline}
\label{sec:pipeline}
\begin{figure*}[htbp]
    \centering
    \includegraphics[width=0.75\linewidth]{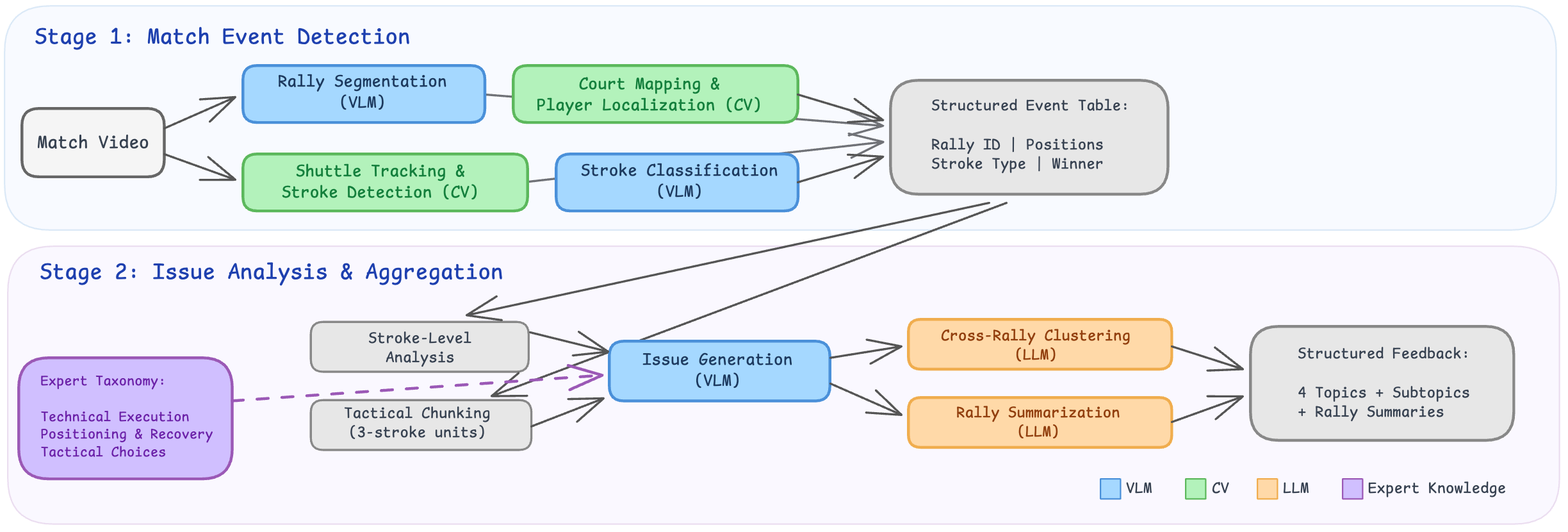}
    \caption{Two-stage computational pipeline that organizes match video into structured events and taxonomy-guided feedback.}
    \Description{A wide flowchart with color-coded modules. Stage 1 (top): Match Video feeds into four modules in a 2x2 grid: Rally Segmentation and Stroke Classification in blue (VLM), Court Mapping and Shuttle Tracking in green (CV), converging into a gray Structured Event Table. Stage 2 (bottom): the event table feeds into two parallel gray boxes (Stroke-Level Analysis and Tactical Chunking), both flowing into a blue Issue Generation (VLM) box. A purple Expert Taxonomy box connects via a dashed arrow. Issue Generation feeds into two orange LLM boxes (Cross-Rally Clustering and Rally Summarization), which converge into a gray Structured Feedback box containing 4 Topics, Subtopics, and Rally Summaries.}
    \label{fig:pipeline}
\end{figure*}

We developed an operational two-stage pipeline that prepares structured inputs for the interface (\autoref{fig:pipeline}). The detection stage extracts match events, including rallies, player positions, and classified strokes. The analysis and aggregation stage maps these events to the expert-informed taxonomy and consolidates them into topic categories and rally-level summaries.

\subsubsection{\textbf{Stage 1: Match Event Detection}}
\label{sec:detection}

In this stage, raw match videos, captured from a monocular fixed high-angle broadcast view, are processed into structured rally and stroke events that serve as input to the feedback generation pipeline (\autoref{fig:pipeline}).

\textbf{Rally segmentation.}
We segment matches into individual rallies using the Doubao-seed-1.6 VLM\footnote{\url{https://console.volcengine.com/ark/region:ark+cn-beijing/model/detail?Id=doubao-seed-1-6}}. The video is divided into one-minute segments at 1\,fps; the VLM identifies rally boundaries, winners, and win reasons. Overlapping clips handle boundary cases. On 109 test rallies, segmentation accuracy was 75\%, with winner/win-reason accuracy of 68\% on successfully segmented rallies.

\textbf{Court mapping and player localization.}
Court boundaries are detected via MonoTrack’s graph-based optimization \cite{monotrack}. Player positions are estimated using a top-down HRNet model \cite{HRnet, mmpose} and projected onto a standardized court plane via homography mapping of ankle midpoints.

\textbf{Shuttle tracking and stroke detection.}
Shuttlecock trajectories are tracked using the WASB model \cite{WASB}. Hitting events are detected by a geometry-based algorithm that identifies peaks in vertical position, velocity changes, and trajectory turns \cite{soloshuttlepose, hsu2024enhancing} (precision 0.71, recall 0.72 on 564 hits).

\textbf{Stroke classification.}
Based on expert recommendations, shots are categorized into 9 types (clear, smash, drop, drive, rush, push, lob, net shot, cross-court net shot) plus 2 serve types. The VLM classifies shot type and handside from 11-frame windows centered on each hit; however, classification accuracy was insufficient and required manual correction. Hitter identification (based on shuttle-arm distance) achieved 77.9\% accuracy.

The resulting structured event table records, for each frame: rally ID, both players’ court positions and zones; for each stroke: hitter, handside, shot type; and for each rally ending: winner and win reason. For the user study, a trained annotator with Level-3 badminton proficiency verified and corrected, when necessary, the downstream stroke-level and rally-level labels, including stroke classifications, rally outcomes, and dependent labels. This review took approximately 25 minutes per 8-minute match after expert alignment, example review, and schema training. Trajectory and player-position traces were not manually reviewed. The reported detection metrics therefore motivate this verification step and limit claims about autonomous raw-video processing.

Different upstream errors affect different parts of the review workflow. Rally-boundary and outcome errors most directly affect the Losing Patterns and Winning Patterns reports and the Rally Explainer. Hit-detection, shot-type, and hitter-identification errors affect stroke-level and tactical-chunk diagnoses. Trajectory and player-position errors, which were not manually reviewed for the study, affect heatmaps and coordinate-level evidence. These dependencies make fully autonomous use and lower-burden verification important directions for future work.

\begin{figure*}[!hbtp]
    \centering
    \includegraphics[width=0.85\linewidth]{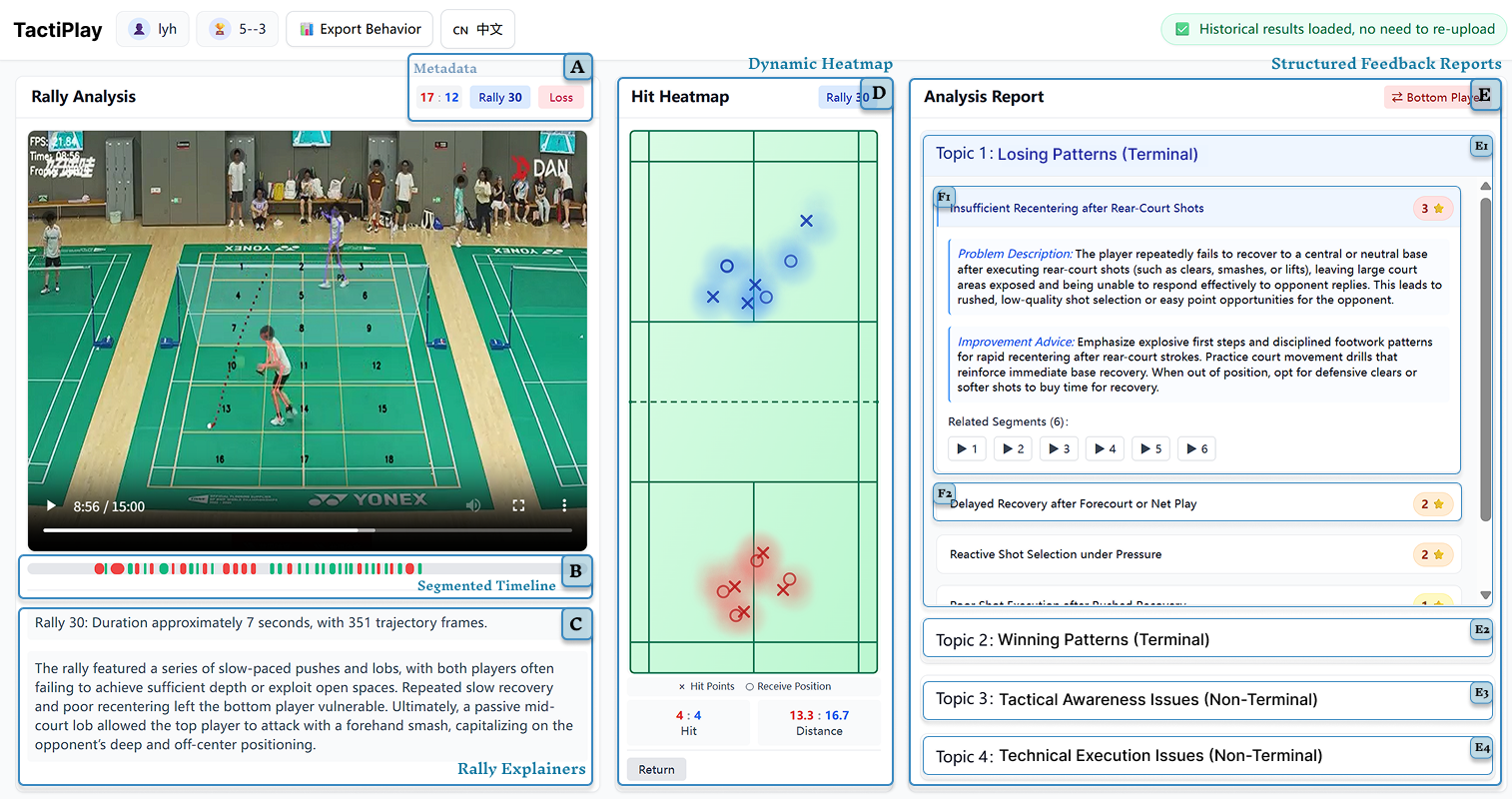}
    \caption{The \name{} interface. (A)~Rally metadata and match progress. (B)~Color-coded rally timeline (green\,=\,win, red\,=\,loss). (C)~Rally Explainer with auto-generated textual summary. (D)~Dynamic court heatmap showing striker ($\times$) and receiver ($\circ$) positions. (E)~Structured feedback report with four topics: Losing Patterns~(E1), Winning Patterns~(E2), Tactical Awareness Issues~(E3), and Technical Execution Issues~(E4). (F)~Subtopic cards: expanded~(F1) with problem description, improvement advice, and linked video segments; collapsed~(F2) showing title and frequency-based star rating.}
    \Description{A screenshot of the TactiPlay interface showing six coordinated panels: (A) match metadata, (B) a color-coded rally timeline, (C) a rally explainer with textual summary, (D) a court heatmap with player position markers, (E) a structured feedback report with four expandable topic categories, and (F) expanded and collapsed subtopic cards with linked video segments.}
    \label{fig:main}
\end{figure*}

\enlargethispage{\baselineskip}
\subsubsection{\textbf{Stage 2: Issue Analysis and Aggregation}}
\label{sec:prepar_system}

We first obtain the feature-augmented stroke sequence and the associated player positions mapped onto the singles court plane 
(\autoref{sec:detection}), which serve as the lowest level of analysis for each rally segment. 
However, individual strokes are often too atomic to capture meaningful tactical patterns. 
To address this, we segment each rally into tactical units, including prefix tactics (one or two opening strokes), 
three-stroke tactical chunks, and terminal tactics (win or loss). 
This design follows prior analyses in racket sports, which identify three-shot subsequences as the minimal tactical unit~\cite{wang2021tac, he2024vistec}. 
Three consecutive strokes form an observable action--response--outcome pattern and thus provide a suitable granularity for tactical analysis.

We used GPT-4.1 as a vision-language model to support multiple analysis tasks in this stage.
For issue generation, each prompt receives both visual and structured input: an annotated hit-frame image showing player skeletons and color-coded shuttle trajectories (distinguishing incoming and outgoing paths), together with the detected event features (stroke types, player positions, shuttle coordinates) and the taxonomy descriptors as a reference frame. The model outputs JSON-formatted issues tagged with category, descriptor, and evidence grounded in court coordinates.
Specifically, the model (1) generates issues at both stroke and tactic levels,
(2) produces rally-level summaries, and (3) clusters scattered issues into coherent \textit{subtopics} and associates them with relevant rally segments.

Building on these tactical units, we adopt a hierarchical descriptor scheme that also aligns with our taxonomy introduced in \autoref{sec:taxonomy}. 
At the \textbf{stroke level}, descriptors capture \textit{Technical Execution} problems (\autoref{fig:taxonomy}), since execution quality is attributable to a single shot.
At the \textbf{tactic level}, descriptors capture \textit{Positioning and Recovery} and \textit{Tactical Choices} (\autoref{fig:taxonomy}), which emerge more meaningfully when analyzed within and across short sequences. 
Following this scheme, we output all detected \textbf{issues} with their category, descriptor, concise evidence grounded in court coordinates or discrete areas, 
and actionable suggestions for improvement.

\textbf{Terminal tactics.} 
We treat the final part of a rally as a terminal tactic, defined as its last three strokes 
(or fewer if the rally ended earlier). 
Unlike regular chunks, terminal tactics must be interpreted together with the rally outcome. 
In this process, we jointly consider both the \textit{tactic-level issues} corresponding to the three-stroke window itself 
and the \textit{stroke-level issues} identified within each of its constituent shots. 
From the loser's perspective, these issues are used to generate candidate explanations for the rally loss
and targeted coaching suggestions.
From the winner's perspective, the same evidence is used to describe how the opponent's observable problems may have created the winning condition,
while also pointing out any potential imperfections on the winner’s side. 
All issues are merged without distinguishing descriptor categories, and aggregated once from the loser’s perspective 
and once from the winner’s perspective. 
We then cluster these aggregated issues into representative \textbf{subtopics}, 
which summarize common losing and winning patterns at the end of rallies 
and form the basis of \textbf{Topic 1: Losing Patterns (Terminal)} and \textbf{Topic 2: Winning Patterns (Terminal)} 
in \autoref{sec:structure_report}. 

\textbf{Non-terminal tactics.} 
Beyond the terminal segments, we also analyze non-terminal tactics to capture how issues accumulate 
throughout the course of a rally. 
These intermediate chunks are process-oriented: instead of being tied directly to rally outcomes, 
they reveal recurring weaknesses such as repeated execution flaws, unstable recovery patterns, 
or consistently reactive shot choices. 
Here, issues are first grouped by descriptor categories 
(\textit{technical execution}, \textit{positioning and recovery}, and \textit{tactical choices}), 
and then aggregated across chunks to reveal systematic weaknesses that unfold during the rally process. 
To further organize these recurring problems into coherent patterns, 
we apply clustering on the aggregated issues to derive fine-grained \textbf{subtopics}. 
These subtopics capture recurring tactical awareness and technical execution problems, 
and serve as the foundation for \textbf{Topic 3: Tactical Awareness Issues (Non-Terminal)} 
and \textbf{Topic 4: Technical Execution Issues (Non-Terminal)} in \autoref{sec:structure_report}. 

\textbf{Rally-level summarization.} 
Finally, we synthesize the outputs of all analysis units within a rally, including prefix tactics (one or two opening strokes), three-stroke tactical chunks, terminal tactics (win or loss), and stroke-level analyses, into a concise textual summary. 
This summary distills the key problems, tactical developments, and decisive moments of the rally into an accessible narrative form. 
It serves as the basis of the \textbf{Rally Explainer panel} (\autoref{fig:main}-C) described in \autoref{sec:rally_review}.
It provides players with an at-a-glance account of what happened in each rally
before they explore more detailed issue categories and feedback in subsequent panels.

In summary, the pipeline organizes detected strokes, positions, and trajectories into tactical chunks, taxonomy-grounded issues, rally summaries, and cross-rally clusters. These outputs feed the interactive modules described next.

\section{\name{}: System Overview}

\name{} is a video-based post-match review system for amateur badminton players. Building on the design requirements from our formative study (\autoref{sec:dr}), it integrates three modules:
\textbf{structured feedback reports} that organize analyses into taxonomy-grounded topics linked to video evidence (DR1, DR2; \autoref{fig:main}-E);
\textbf{rally-centered review} that treats rallies as the natural unit of reflection, distinguishing outcome-oriented from process-oriented analyses (DR3; \autoref{fig:main}-A--D);
and \textbf{dynamic court visualizations} that project player positions and stroke contexts onto a standardized court synchronized with video playback (\autoref{fig:main}-D).
Together, these modules transform match events into structured narratives that help players understand \textit{what happened}, which observable patterns contributed to the outcome, and \textit{how to improve}.

\subsection{Structured Feedback Reports (DR1, DR2)}
\label{sec:structure_report}

The structured report (\autoref{fig:main}-E) organizes pipeline outputs into four topics (\autoref{sec:prepar_system}). Two \textbf{outcome-level} topics capture terminal tactics: \textit{Losing Patterns}~(E1) identifies recurrent ways rallies were lost, while \textit{Winning Patterns}~(E2) highlights successful strategies. Two \textbf{process-level} topics capture non-terminal issues: \textit{Tactical Awareness Issues}~(E3) surfaces decision-making problems such as missed opportunities, and \textit{Technical Execution Issues}~(E4) pinpoints execution flaws that accumulate over time. This separation of tactical and technical dimensions addresses \textbf{DR2}.

Each topic is decomposed into \textit{subtopics}, recurring patterns obtained by clustering aggregated issues (\autoref{sec:prepar_system}), presented in descending frequency. Subtopic cards display a problem description with improvement suggestions, or for Topic~2, successful strategies with refinement notes, together with all associated video segments. Pressing the play button replays the corresponding snippet, grounding every piece of feedback in observable performance (\textbf{DR1}).

\subsection{Rally-Centered Review (DR3)}
\label{sec:rally_review}

A segmented timeline (\autoref{fig:main}-B) divides the match into rallies color-coded by outcome. Selecting a rally opens synchronized video playback with contextual metadata: rally ID, score, outcome, duration, and stroke count. Beneath the video, the \textbf{Rally Explainer} (\autoref{fig:main}-C) provides an auto-generated summary describing the rally's flow, momentum shifts, and terminal cause, enabling players to grasp what happened without replaying every stroke. These elements operationalize rallies as the natural unit of reflection (\textbf{DR3}).

\subsection{Dynamic Court Visualization}

The dynamic heatmap (\autoref{fig:main}-D) projects CV detections onto a standardized singles court. At each stroke, a cross marks the striker's position and a circle marks the receiver's, enabling assessment of spatial tendencies and positional vulnerabilities. A density overlay aggregates positions across all rallies or a selected rally to reveal court coverage patterns. During playback, the video panel overlays synchronized court segmentation, player skeletons, and shuttle trajectories, providing time-aligned spatial context.

\section{User Study}

We conducted a controlled within-subjects experiment to evaluate how amateur players engage with \name{} and how its integrated review workflow affects tactical reflection. Both conditions used match packages prepared with the same upstream event-detection pipeline, input schema, and manual-verification protocol, including checks of stroke-level and rally-level labels, and both report generators used GPT-4.1. Each participant reviewed two different matches across conditions, with match-condition assignment and system order counterbalanced; thus, data preparation rather than match content was held constant. The study tests how the two review systems organize and link reviewed match evidence, not autonomous raw-video processing.
Each participant played and recorded two matches from a fixed broadcast-view camera angle.
We compared subjective ratings, reflection quality, and qualitative feedback across conditions.

\subsection{Baseline System}
\label{sec:baseline}
\begin{figure}[htbp]
    \centering
     \includegraphics[width=1\linewidth]{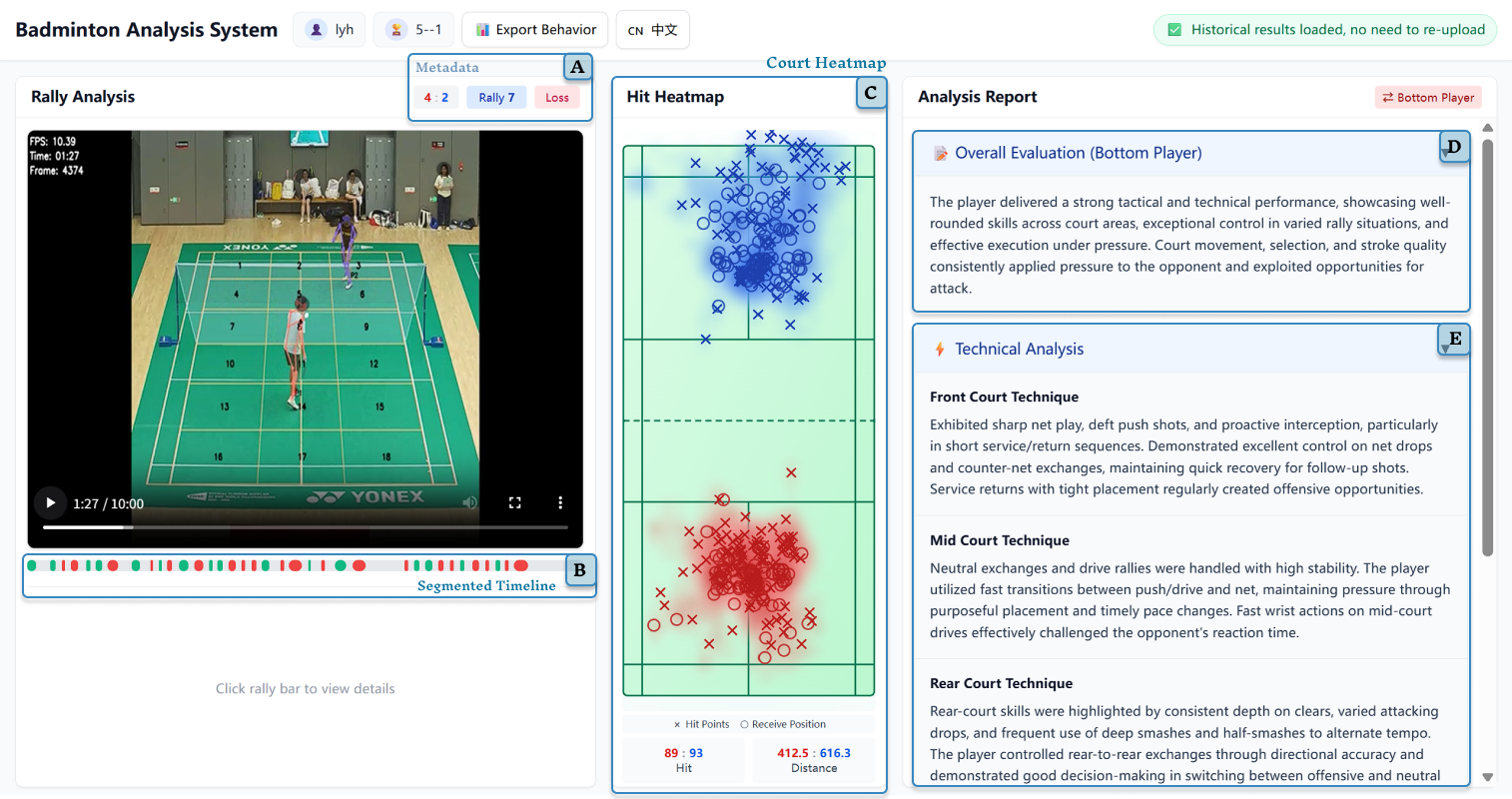}
    \caption{Baseline report-and-statistics system: (A) rally metadata, (B) clickable timeline with basic rally-to-video playback but no rally-specific explanation, (C) global heatmap, (D) overall evaluation, and (E) five-dimensional technical analysis. Report items are not video-linked, and rally selection does not update the report or heatmap.}
    \Description{A screenshot of the baseline system interface showing five panels: (A) rally metadata, (B) a clickable color-coded rally timeline used to play a selected rally without a rally-specific explanation, (C) a static court heatmap with aggregated shot distributions that does not update with rally selection, (D) an overall evaluation paragraph, and (E) a technical analysis panel with five dimensions of play. The report and heatmap are not coordinated with the timeline, and report items do not link to rally clips.}
    \label{fig:baseline}
\end{figure}

We designed the baseline as a representative report-and-statistics system inspired by commercial products such as \textit{Shanyu AI}, not as a state-of-the-art benchmark or component ablation. It generated a fixed summary report from its assigned match package (\autoref{fig:baseline}), whereas \name{} used taxonomy-guided issue generation, rally-level organization, and evidence-linked interaction.
The baseline report comprised an \textbf{Overall Evaluation} (D) and a \textbf{Technical Analysis} (E) covering front court, mid court, rear court, defense, and attack-defense transitions. Its timeline (B) supported basic rally-to-video navigation, but selection did not update the heatmap or report or provide a rally-specific explanation. The heatmap (C) remained a global aggregate without rally-level filtering or synchronized trajectory replay.
Thus, the baseline combined a fixed report and aggregate statistics with basic video navigation. In contrast, \name{} provided a taxonomy-organized report, subtopic-linked clips, rally-specific explanations through the Rally Explainer, and rally-filtered heatmaps synchronized with selected video evidence. The comparison supports the integrated workflow as a whole, not any single component.

\subsection{Participants}
We recruited 16 amateur badminton players (10 male, 6 female; aged 20--27, $M = 23.87$, $SD = 3.66$) from university clubs and recreational groups via social media and word of mouth.
Self-reported skill levels ranged from Level~2 to Level~4 on the \textit{BadmintonCN} nine-level scale (\autoref{sec:dr}), matching our target population: novice-to-lower-intermediate recreational players who possess basic technical skills and play regularly but lack systematic tactical training or professional coaching.
Participants had 1--10 years of playing experience ($M = 4.13$, $SD = 3.11$) and played at least once per week.

\subsection{Procedure and Tasks}

The study received institutional IRB approval; all participants gave informed consent.
To ground the reflection tasks in authentic playing experiences, we invited participants one week before the main session to play two matches in a gymnasium equipped with an overhead camera.
Each player was randomly paired with two other participants and took a different side in each match. Matches lasted 8.03 $\pm$ 1.85 minutes on average.
Both recordings from each participant were processed with the same upstream event-detection pipeline (\autoref{sec:detection}) and input schema. A trained Level-3 badminton annotator applied the same manual-verification protocol to stroke-level and rally-level labels ($\sim$25 minutes per 8-minute match). The two match-specific packages were assigned to the systems in counterbalanced order.

Each main session lasted approximately 80--100 minutes and followed four steps.
\textbf{(1)~Introduction.} Participants were welcomed, briefed on the study objectives, and signed informed consent.
\par\noindent\textbf{(2)~First condition.} Participants reviewed one of their own matches using the baseline or \name{} at their own pace, identifying key problems and areas for improvement. Afterward, they completed (a)~a Likert-scale questionnaire on subjective perceptions of the review process, reflection outcomes, and perceived workload, (b)~a reflection task in which they listed the key problems and improvement points they identified (later categorized and scored for reflection quality), and (c)~a semi-structured interview covering which features and feedback modalities were most helpful, whether the feedback aligned with perceived performance, and suggestions for improvement.
\par\noindent\textbf{(3)~Second condition.} Participants repeated the same procedure with the other system and a different match video. System order and match-side assignments were counterbalanced using a $2 \times 2$ Latin square design.
\par\noindent\textbf{(4)~Comparison.} Participants compared the two systems in a cross-condition interview covering helpfulness, differences in their reflection process, and support for identifying detailed problems and tactical opportunities.

\subsection{Measurements}

We evaluated the two review conditions from three perspectives:

\subsubsection{Reflection Quality Assessment.}
After each review session, participants listed the key problems they identified and corresponding improvement suggestions. We assessed these reflection entries on \textit{frequency} and \textit{quality} \cite{wiggins2012seven, leijen2012determine, bialecki2023esport}, categorizing each entry using the performance taxonomy (\autoref{sec:taxonomy}): \textit{Technical Execution}, \textit{Positioning and Recovery}, and \textit{Tactical Choices}. An entry could receive labels from multiple taxonomy domains when it addressed more than one aspect of play.
Two professional badminton players rated each entry on three 7-point dimensions while checking it against the corresponding match footage. We removed condition labels from all entries before rating. The rubric used three badminton-specific dimensions. \textbf{Concreteness} ranged from a vague statement such as ``my defense was bad'' to a reflection that identified a specific rally, shot sequence, court area, or decision context \cite{leijen2012determine}. \textbf{Actionability} ranged from general advice such as ``play better'' to a specific adjustment in recovery direction, shot choice, or follow-up action \cite{wiggins2012seven}. \textbf{Appropriateness} ranged from a proposed fix that did not match the visible problem to one that logically addressed the problem shown in the footage \cite{leijen2012determine}.
The raters first calibrated on a shared subset, then independently categorized and scored all entries. They resolved categorization disagreements through discussion and averaged the quality scores. Inter-rater reliability was high (Concreteness ICC\,=\,.875; Actionability ICC\,=\,.872; Appropriateness ICC\,=\,.906; all $p < .001$).

\subsubsection{Subjective Ratings.}
After each condition, participants rated their experience on 7-point Likert items (1 = Not at all, 7 = Very much). Rather than adopting a single validated instrument, we designed targeted single-item probes organized around three aspects of the review experience:
\textit{Participant-perceived feedback quality} (helpfulness, actionability, credibility) captures participants' judgments of whether the system output was useful, actionable, and believable \cite{wiggins2012seven, weidlich2025highly}.
\textit{System usefulness and review process} (efficiency, satisfaction, ease of use, functional completeness and integration, intention to continue use) captures how the system shapes practical review. We assessed completeness and integration jointly because \name{} combines report, video, and visualization as a cohesive workflow rather than as independent modules.
\textit{Reflection support} (performance understanding, problem identification, tactical awareness) captures participants' perceptions of whether the review helped them understand their performance \cite{groom-2011-video-pa, kleygrewe2024changing}. These subjective constructs are distinct from the expert-rated quality of participants' written reflections. Neither measure constitutes objective per-item validation of every generated tactical diagnosis.
Three exploratory workload items (mental demand, effort, frustration), adapted from standard workload dimensions \cite{hart1988development}, captured perceived cognitive burden.

\subsubsection{Interaction Logs and Qualitative Insights.}
\label{sec:measure-interaction}
System logs captured usage time, rally replay counts, navigation sequences, and subtopic click rates (the latter for \name{} only). Two authors open-coded transcripts from the two post-condition interviews and the exit interview to identify recurring themes across conditions. For the interaction-strategy comparison, we interpreted these coded accounts together with the observed order of report, video, and heatmap use in the logs.

\section{Results}
\label{sec:results}

We report quantitative comparisons using Wilcoxon signed-rank tests \cite{wilcoxon1945individual} for paired data and Mann--Whitney U tests \cite{mann1947test} for independent samples, complemented by the interview coding and interaction-log analysis described in \autoref{sec:measure-interaction}.

\begin{figure*}[!tb]
    \centering
    \includegraphics[width=\linewidth]{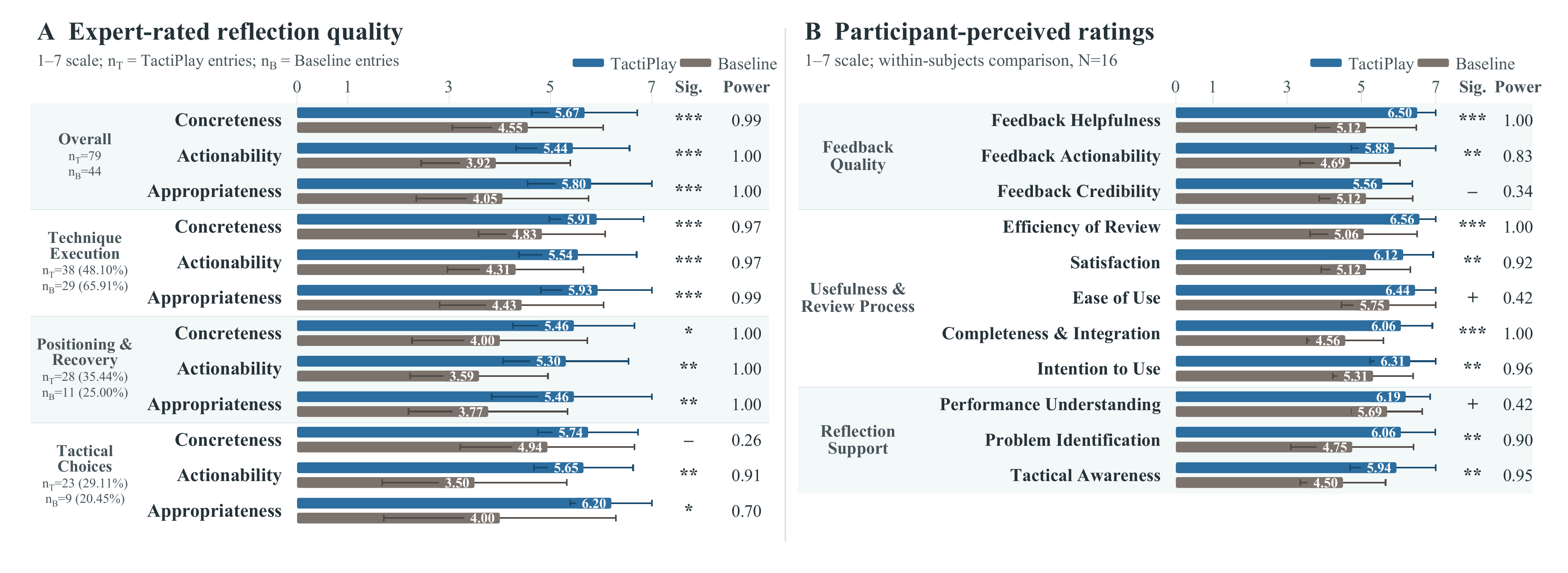}
    \caption{Comparison of \name{} and baseline evaluation results. Bars show means for 1--7 ratings from a zero baseline; whiskers show $\pm 1$ SD, clipped at the scale bounds. \textit{Part A}: Expert-rated reflection quality; $n_T$ and $n_B$ denote condition-level entry counts. Domain coding was multi-label, so percentages need not sum to 100\%. \textit{Part B}: Participant-perceived ratings (within-subjects, $N=16$). Significance: $-$ $p>.10$; $+$ $.05<p<.10$; $*$ $p<.05$; $**$ $p<.01$; $***$ $p<.001$. Rightmost values report observed power.}
    \Description{Two side-by-side grouped horizontal bar-chart panels compare TactiPlay with the baseline. Each panel includes its own legend. Part A reports expert-rated reflection quality for 12 combinations of reflection category and quality dimension, including condition-specific entry counts. Reflection-domain coding was multi-label, so category percentages can sum above 100 percent. Part B reports 11 participant-perceived measures grouped into feedback quality, usefulness and review process, and reflection support. Muted blue bars represent TactiPlay and muted warm-gray bars represent the baseline; white labels give exact means on 1--7 scales, darker same-color whiskers show one standard deviation, and right-hand columns report significance and observed power. TactiPlay has higher means on nearly all measures; feedback credibility is the only Part B measure without a significant or marginal difference.}
    \label{fig:evaluation_results}
\end{figure*}

\subsection{Reflection Outcomes}
\label{sec:reflection-results}

We first report the quality of reflections participants produced after each session, assessed by two professional badminton players (\autoref{fig:evaluation_results}, Part~A). This was our primary measure of reflective outcomes.

\textbf{Frequency.}
\name{} elicited nearly twice as many reflection entries as the baseline (79 vs.\ 44). Because entries could receive multiple domain labels, percentages use all entries in each condition as the denominator and need not sum to 100\%. Baseline reflections concentrated on technique execution (65.91\%), with fewer entries on positioning/recovery (25.00\%) and tactical choices (20.45\%). \name{} entries were less concentrated on technical execution and more broadly distributed across positioning/recovery and tactical choices. P3 attributed this shift to the layered feedback structure: \textit{``It not only covered how well I executed a shot, but also addressed my positioning choices and shot selection, so I also considered positioning and tactical decisions.''}

\textbf{Quality.}
\name{} significantly outperformed the baseline on all three quality dimensions overall.
\textit{Concreteness} improved significantly for technique execution and positioning/recovery, though not for tactical choices. Linked clips anchored reflections in concrete rallies, whereas baseline feedback was often too vague to act upon. As P5 noted, the baseline merely stated that transitions between offense and defense were \textit{``worth noting''} without specifying the actual issue.
\textit{Actionability} ($M=5.44$, $SD=1.12$) showed significant advantages across all categories. Under the baseline, reflections stopped at describing problems. P16 noted: \textit{``There were hardly any concrete suggestions, so I couldn't turn them into a training plan.''} With \name{}, participants formulated concrete improvement plans. P7 explained: \textit{``I realized I relied too much on straight shots, so I need to play more cross-court and recover earlier.''}
\textit{Appropriateness} ($M=5.80$, $SD=1.25$) was likewise significantly higher across all categories, with \name{}'s reflections showing tighter alignment between diagnosed problems and proposed fixes. P16 noted: \textit{``It pointed out that my late recovery cost me the initiative, and suggested that after a diagonal clear I should immediately return to the center---a fix that directly echoed the problem itself.''}

\subsection{Subjective Experience}
\label{sec:subjective-results}

Part~B of \autoref{fig:evaluation_results} reports participant perceptions rather than objective feedback correctness. Of 11 subjective items, eight favored \name{} significantly, two showed marginal trends, and one (feedback credibility) showed no significant difference. Three exploratory workload items showed no statistically significant differences between conditions; participant comments suggested that reduced manual scanning was offset by report volume and technical language.

\textbf{Participant-perceived feedback quality.}
Participants rated \textit{helpfulness} and \textit{actionability} significantly higher for \name{}. They valued that each subtopic linked to video segments across rallies, revealing recurring problems translatable into training goals (P16: \textit{``The video examples under each point were very specific, so I knew exactly what to work on next''}). \textit{Credibility} ratings were slightly higher but not significantly different. The baseline's vagueness paradoxically felt ``realistic'' to some participants, while \name{}'s detail made small analytical mismatches immediately visible (P3: \textit{``Sometimes the system said I was intercepted, but in the clip the shuttle was actually out. That made me question the analysis''}).

\textbf{System usefulness and review process.}
Participants rated \textit{review efficiency} significantly higher for \name{}: linked clips reduced manual scene-searching (P1: \textit{``\name{} saved me time\ldots I could immediately watch dense clusters of related clips''}). \textit{Satisfaction} ratings were also significantly higher, with participants describing the feedback as \textit{``very intuitive''} (P1), \textit{``thorough and detailed''} (P4), and \textit{``direct rather than abstract''} (P16), compared to the baseline which was perceived as ``polite but vague.'' Ratings of \textit{completeness and integration} and \textit{intention to use} were likewise significantly higher, with participants praising the seamless linkage between report, video, and heatmap.

\textbf{Reflection support.}
Participants rated support for \textit{problem identification} and \textit{tactical awareness} significantly higher under \name{}. Grouped exemplars made weaknesses salient and verifiable, and linked rally replays turned abstract descriptions into concrete decision-making insights (P1: \textit{``Sometimes the tactical text felt vague, but once I clicked the linked clips, I immediately understood what it meant''}). Baseline feedback was perceived as broad, praise-oriented, and loosely coupled with evidence (P6, P9, P13). \textit{Performance understanding} ratings were slightly higher for \name{} but not significantly different, as the baseline already provided a sufficient match overview.

\subsection{Interaction Patterns}
\label{sec:interaction-patterns}

\begin{figure*}[!tb]
    \centering
    \includegraphics[width=0.85\linewidth]{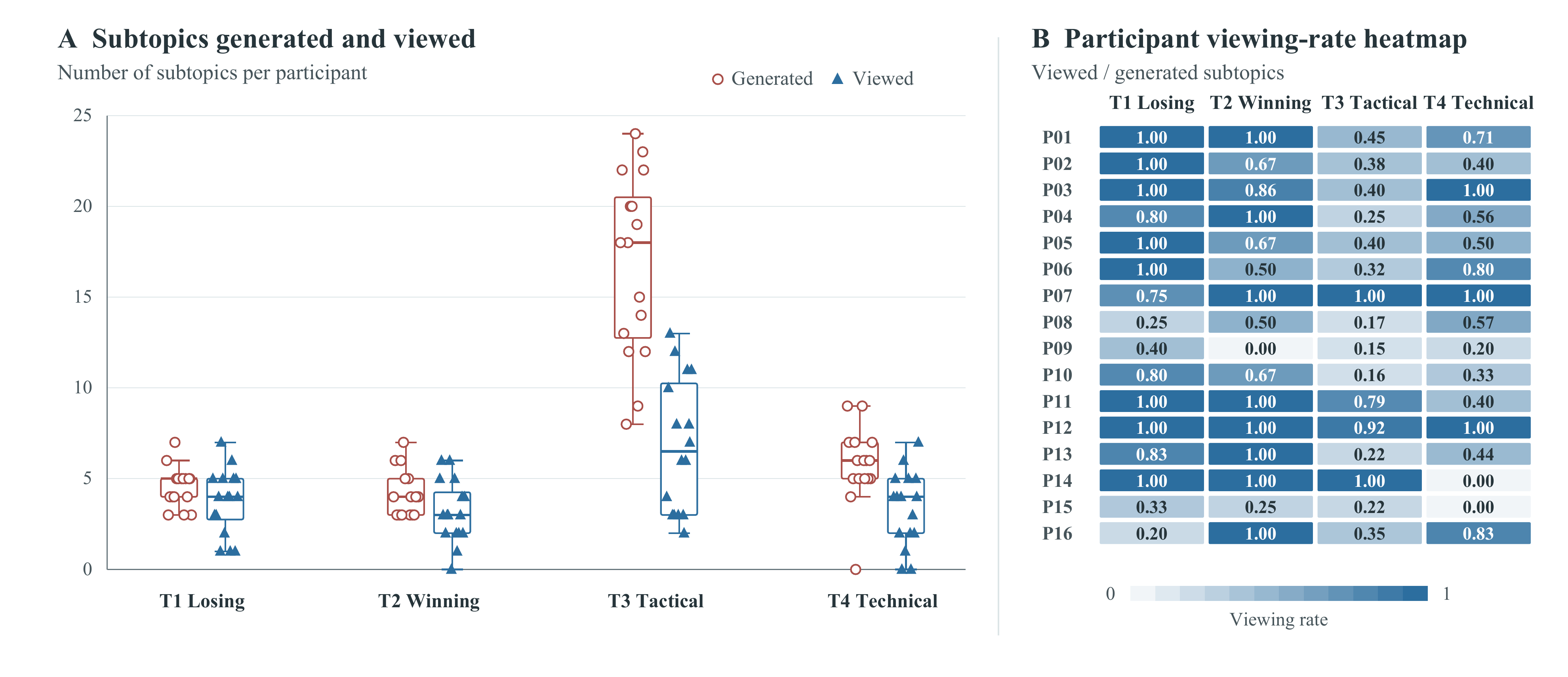}
    \caption{Interaction logs within \name{}. \textit{(A)} Generated and viewed subtopics by participant and feedback topic; points show participants and boxplots summarize distributions. \textit{(B)} Per-participant viewing rate (viewed/generated, capped at 1.0) across the four topics.}
    \Description{Two-panel visualization of TactiPlay interaction logs. Panel A shows participant-level generated and viewed subtopic counts for four feedback topics. Hollow muted-red circles indicate generated counts, filled blue triangles indicate viewed counts, and paired boxplots summarize each distribution. Panel B is a 16-by-4 heatmap of viewed/generated rates, capped at 1.0, for participants P01 through P16 across Losing Patterns, Winning Patterns, Tactical Awareness, and Technical Execution. Every cell displays its numeric rate from 0.00 to 1.00, with darker blue indicating a higher rate.}
    \label{fig:topic_interaction}
\end{figure*}

\autoref{fig:topic_interaction} summarizes generated and viewed subtopics within \name{}. Interviews and coded review sequences revealed a condition-level shift in review strategy: with \name{}, 12 of 16 participants used a report-driven strategy and four used a video-driven strategy; under the baseline, only six were report-driven, while 10 relied on video-driven manual search.

\textbf{Report-driven} (12/16 participants): players started from the structured report, formed hypotheses about their weaknesses, and validated them through linked clips and the heatmap, forming an iterative verification loop. P7: \textit{``I carefully read each item, then clicked the linked rally to watch, and used the heatmap to confirm the positioning issue.''}

\textbf{Video-driven} (4/16 participants): players relied on direct video inspection via the timeline and heatmap, consulting the report afterward. P9: \textit{``I first used the timeline to filter out lost rallies, then checked the heatmap, and only afterward looked at the text analysis.''}

Under the baseline, report-driven users found verification effortful because the report did not link directly to video evidence. As P12 noted: \textit{``When it said my net play was unstable, I had already forgotten which rally it was.''} This reversal from predominantly video-driven search in the baseline to predominantly report-to-evidence review in \name{} is consistent with a change in how participants moved among reports, video, and spatial evidence.

\section{Discussion}

This work addresses the challenge amateur badminton players face in deriving tactical insights from match recordings.
Participants rated the review process as more efficient with \name{}, and their written reflections shifted from vague impressions toward more concrete, actionable, and appropriate tactical reasoning.
We first discuss how the formative design requirements (DR1--DR3) were realized, then present design implications from our evaluation.

\subsection{Design Implications}
Our formative study identified three design requirements: DR1 (feedback linked to concrete match footage), DR2 (attention to tactical awareness), and DR3 (rally-level entry points for reflection). \name{} operationalizes each through its core modules: video-anchored subtopic clips (DR1), the taxonomy-grounded four-topic structure (DR2), and the rally-centered timeline with Rally Explainer (DR3). Building on the evaluation, we identify three further design implications for future systems.
The observed strategy reversal (\autoref{sec:interaction-patterns}) provides the empirical starting point: 12 of 16 participants used a report-to-evidence loop with \name{}, whereas 10 of 16 relied on video-driven manual search with the baseline. The implications below distinguish observations supported by this comparison from directions motivated by the study's limitations.

\subsubsection{\textbf{Explaining Tactical Opportunities from Observable Play}}
The report-to-evidence loop shows that participants used linked clips and heatmaps to interpret and verify structured feedback. This interaction pattern supports explanations that connect factual events to observable tactical opportunities, such as a weak opponent return that created an attacking chance or a slow recovery that exposed open court space \cite{wu2021tacticflow, he2025finebadminton}. These explanations should remain grounded in shot sequences, court positions, opponent responses, and rally outcomes rather than claim access to a player's private mental state.
\textbf{Design Implication 1 (DI1): Explain tactical opportunities with observable match evidence.} Future systems can identify when visible play created, preserved, or closed an opportunity and link that explanation to the relevant sequence. Explicit review controls can still support different levels of detail, but personalization should rely on user-provided goals and observable behavior rather than inferred private intentions.

\subsubsection{\textbf{Towards Actionable Alternatives}}
Although \name{} links textual suggestions to video clips, participants still asked to see \textit{how} an improvement could be performed in situ, such as adjusted footwork or an alternative return trajectory. This request extends the observed value of evidence linking from showing the problem to situating a possible response. Prior work has shown that counterfactual visualizations improve causal understanding \cite{kaul2021improving}, and situated annotations increase feedback clarity in sports \cite{lin2025sportsbuddy, zhu2022sporthesia}.
\textbf{Design Implication 2 (DI2): Visualize alternative actions in the original play context.} Future systems can overlay suggested movement paths or return trajectories on court maps or video frames, allowing players to compare the observed action with a concrete alternative.

\subsubsection{\textbf{Towards Longitudinal Tracking}}
Our study reviewed one match per condition and therefore cannot distinguish isolated mistakes from persistent weaknesses. We derive the next implication from this study boundary rather than from a direct participant finding. Prior work shows how longitudinal and cross-session views can reveal sustained performance trends \cite{wu2017ittvis, perin2013soccerstories, till2022optimising}.
\textbf{Design Implication 3 (DI3): Support longitudinal tracking and cross-match comparison.} Future systems can visualize how issues recur or diminish across games, map tactical diversity over time, and compare performance against different opponents. Such views would help players distinguish one-off errors from systematic habits.

\subsection{Broader Implications}
Our structured, taxonomy-grounded feedback approach could extend to other racket sports (tennis, table tennis, squash), where similar rally-based tactical structures exist \cite{memmert2017current, yin2025construction, paulauskas2025padel}. Cross-sport transfer would still require sport-specific validation of the taxonomy and event representations rather than direct reuse of badminton categories. In coaching contexts, AI can transform underutilized match recordings into structured reflection materials \cite{bridgeman2024using, jud2025ai}, acting as an ``assistant coach’’ that compiles key rallies for coaches to review efficiently \cite{wang2024tacticai}.

However, AI feedback must align with coaching philosophy to avoid confusion \cite{naughton2024challenges}, and interpretability, such as linking diagnoses to video evidence, is essential for building trust \cite{hammes2022artificial}. Our findings also highlight the importance of adaptive feedback depth: some participants preferred concise cues, while others valued detailed tactical chains, suggesting that future systems should tailor complexity to users’ needs \cite{gao2025role, exel2024precision, pashaie2024unlocking}.

At a broader level, AI coaching tools have the potential to connect recreational training, youth development, and professional practice, aligning with Long-Term Athlete Development (LTAD) principles \cite{till2022optimising, naughton2024challenges}.

\subsection{Limitations and Future Work}
First, our participants were novice-to-intermediate amateur singles players (N=16) aged 20--27, limiting generalizability across ages, skill levels, and doubles play. Feedback needs and reflection practices vary across ages and expertise \cite{ericsson2006influence}, while doubles introduces different tactical demands \cite{gawin2015competition}. Future studies should recruit broader samples and test whether rally reconstruction, tactical chunking, and taxonomy-based categorization transfer to doubles and other racket sports or require alternative, customizable views.
Second, our subjective items targeted specific aspects of the review experience rather than validated psychological constructs. Although informed by established assessment dimensions, the custom items lacked formal psychometric validation; future work should use standardized instruments or validate new scales.
Third, we evaluated reflection rather than subsequent match performance, so whether the observed insights changed behavior remains unknown. Timely feedback can support actionable learning \cite{baca-2006-rapid-feedback}, but our post-match design may reduce immediacy. Longitudinal studies should compare post-match and more immediate feedback and measure whether either produces sustained tactical adaptation.
Fourth, our study evaluated the integrated review workflow using manually verified stroke-level and rally-level labels; trajectory and player-position traces were not manually reviewed. Verification took approximately 25 minutes per 8-minute match, limiting immediate deployment and motivating more reliable event detection and field testing \cite{chaudhury2019unsupervised, arbues2019single}. Without verification, rally-boundary and outcome errors can affect report topics and Rally Explainers; hit-detection, shot-type, and hitter-identification errors can affect stroke- and chunk-level diagnoses; and trajectory or player-position errors can affect heatmaps and coordinate-level evidence. We also did not systematically validate the objective correctness of each generated tactical diagnosis. Per-item expert validation remains necessary before autonomous deployment.
Fifth, the comparison supports the integrated workflow rather than independent component effects. Both conditions used match packages prepared with the same upstream event-detection pipeline, input schema, and manual-verification protocol, and both report generators used GPT-4.1. However, each participant reviewed two different matches, with match-condition assignment counterbalanced. The conditions also differed in prompt structure, taxonomy organization, rally summaries, cross-rally clustering, video linking, interaction, and potentially output length. Component-level and text-length-controlled ablations remain future work.

\section{Conclusion}

We presented \name{}, an interactive system that instantiates an expert-taxonomy-guided, rally-level, video-anchored review workflow designed to help amateur badminton players move from broad impressions to structured tactical reflection. A formative study (N=8) and an expert-derived performance taxonomy informed its analytical pipeline and interactive review interface.

In a within-subjects evaluation (N=16), \name{} elicited more frequent, concrete, actionable, and appropriate reflections than a report-and-statistics baseline. Participants also reflected more broadly on positioning and tactical choices, and 12 of 16 used a report-to-evidence strategy that connected structured feedback to linked clips and court visualizations. These findings support the integrated review workflow under reviewed stroke-level and rally-level labels; they do not establish autonomous raw-video analysis or objective correctness for every generated diagnosis. Overall, organizing reviewed match evidence around taxonomy-guided issues, rallies, and inspectable video links can support amateur players' tactical reflection. Future explanations should connect factual events to observable tactical opportunities without inferring players' private intentions.

\begin{acks}
This project is supported by the HKUST Sports Science and Technology Research Grant (Grant No.: SSTRG24EG07).
\end{acks}

\bibliographystyle{ACM-Reference-Format}
\bibliography{reference}

\end{document}